\documentclass[aps,prb,twocolumn,showpacs]{revtex4}
\usepackage{color}
\usepackage{graphicx}
\usepackage{cancel}
\usepackage{dsfont}

\begin{document}

\title{Robust to impurity-scattering spin Hall effect in two-dimensional electron gas}
\author{V. K. Dugaev$^{1,2}$, M. Inglot$^1$, E. Ya. Sherman$^{3,4}$, and J. Barna\'s$%
^{5,*}$} 
\affiliation{$^1$Department of Physics, Rzesz\'ow
University of Technology,
Powsta\'nc\'ow Warszawy 6, 35-959 Rzesz\'ow, Poland\\
$^2$Department of Physics and CFIF, Instituto Superior T\'ecnico,
Av. Rovisco Pais, 1049-001 Lisbon, Portugal\\
$^3$Department of Physical Chemistry, Universidad del Pa\'is Vasco UPV-EHU, 48080 Bilbao, Spain\\
$^4$Basque Foundation for Science IKERBASQUE, 48011, Bilbao, Spain\\
$^5$Institute of Molecular Physics, Polish Academy of Sciences,
Smoluchowskiego 17, 60-179 Pozna\'n, Poland}
\date{\today }

\begin{abstract}
We propose a mechanism of spin Hall effect in two-dimensional
electron gas with spatially random Rashba spin-orbit interaction.
The calculations based on the Kubo formalism and kinetic equation
show that in contrast to the constant spin-orbit coupling,
spin Hall conductivity in the random spin-orbit
field is not totally suppressed by the potential impurity scattering.
Even if the regular contribution is removed by the vertex corrections,
the terms we consider, remain. 
Therefore, the intrinsic spin-Hall effect exists being,
however, non-universal.
\end{abstract}


\maketitle

Spin currents are believed to be of great importance for future
spin electronics \cite{reviews}, as they offer the possibility of
nonmagnetic manipulation of magnetic moments. Generally, spin
currents are associated with charge currents, and can be generated
by various methods, such as for instance by electric field in
magnetic systems or circularly polarized light in nonmagnetic
semiconductors. Of particular interest, however, are pure spin
currents, where the flow of spins is not accompanied by any
electric current. Search for generation techniques of pure spin
currents, especially in nonmagnetic semiconducting systems, is of
high interest both for fundamental and applied physics. One of the
possibilities of producing pure spin currents relies on the spin
Hall effect (SHE) in nonmagnetic semiconductors with various types
of spin-orbit (SO) interaction, where uniform electric field
causes a transverse spin rather than charge current.

The existence of SHE in semiconductors with impurities has been
predicted by Dyakonov and Perel' \cite{dyakonov71}. Since then
several mechanisms of SHE have been proposed
\cite{hirsch99,murakami03,sinova04}, and the effect has been
observed in a number of experiments \cite{kato04,wunderlich05}. It
is generally believed that two extrinsic mechanisms related to the
SO-dependent scattering by impurities, i.e. side jump and skew
scattering \cite{smit58,berger70,crepieux01}, can be responsible
for the SHE in metallic and semiconducting materials. In addition,
a lot of discussions in recent literature concerned the
possibility of  SHE due to intrinsic SO interaction in
disorder-free systems. An extensively studied example of such a
system is a two-dimensional electron gas with constant Rashba SO
interaction \cite{bychkov84,sinova04}, leading
to the momentum-dependent spin splitting of electron states.
The theoretical efforts were especially focused on the possibility of
equilibrium spin currents \cite{RashbaESC,Tokatly08} and universal
SHE \cite{sinova04} independent on the SO coupling strength.
However, it turned out that the role of impurities is crucial for
this mechanism \cite{engel07}. It has been shown that even in the
limit of a very weak spin-independent disorder, the potential
scattering from impurities suppresses the SHE completely
\cite{EngelRashba,ShytovHalperin,inoue04,dimitrova05,KhaetskiiSH,
hankiewicz08,tarasenko06}.
In the case of random Rashba field without impurity scattering,
the SHE can be also nonzero, as shown numerically
within tight-binding model for a finite-size system \cite{moca08}.

Here we show that the intrinsic SHE does exist,
paradoxically, in relatively dirty systems, where the SO coupling appears locally,
but vanishes on average. Such a random spin dynamics is common in
symmetric semiconductor quantum wells (QWs) \cite{sherman03}, such
as Si/SiGe \cite{Golub04} and GaAs/AlGaAs QWs grown along the
[110] axis \cite{Oestreich}. Moreover, we show that impurities
play here the role less important than in the case of uniform SO
interaction -- the spin Hall conductivity does not vanish in the
limit of small impurity density.
This behavior is qualitatively
different from that for constant Rashba SO interaction.

To describe the model under consideration we assume Hamiltonian of
electrons moving in the ${\mathbf r}=(x,y)$ plane with random Rashba SO interaction,  $%
\hat{H}=\hat{H}_{0}+\hat{H}_{\rm so}$ (in the following we use units with $\hbar
\equiv 1$), with
\begin{eqnarray}
&&\hat{H}_{0}=-\frac{\nabla ^{2}}{2m}, \label{Hamiltonian1}
\\
&&\hat{H}_{\rm so}=-\frac{i}{2}\sigma _{x}\left\{ \nabla _{y},\,\lambda (\mathbf{%
r})\right\} +\frac{i}{2}\sigma _{y}\left\{ \nabla _{x},\,\lambda (\mathbf{r}%
)\right\} ,  \label{Hamiltonian2}
\end{eqnarray}
where $\nabla _{i}=\partial _{i}-eA_{i}/c$, $\mathbf{A}$ is the
vector potential of external field, $e$ and $m$ is the electron charge and
effective mass, respectively,
and $\sigma_{a}$ are the Pauli matrices ($a=x,y,z$). The curly brackets $\left\{ ...\right\} $ stand
for the anticommutator of the appropriate operators to ensure the
Hermitian form of the Hamiltonian. The random coupling parameter $\lambda (%
\mathbf{r})$ vanishes on average, $\left\langle \lambda
(\mathbf{r})\right\rangle
=0$, while the correlator $C_{\lambda \lambda }(\mathbf{r}-\mathbf{r^{\prime }%
})\equiv \left\langle \lambda (\mathbf{r})\lambda (\mathbf{r^{\prime }}%
)\right\rangle =\left\langle \lambda ^{2}\right\rangle F(\mathbf{r}-\mathbf{%
r^{\prime }})$, with all higher correlators reduced to the
second-order one for the Gaussian fluctuations of $\lambda
(\mathbf{r})$.

The spin current operator has the following form \cite{Shi}:
\begin{equation}
\hat{\jmath}_{i}^{a}=\frac{1}{4e}\,\{\hat{\jmath}_{i},\,\sigma
_{a}\},  \label{jia}
\end{equation}
where $\hat{\jmath}_{i}=-c\,(\partial \hat{H}/\partial A_{i})$ is
the  $i$-th component of the current operator ($i=x,y,z$). We consider
in-plane electric field $\mathbf{E}$, and calculate the total spin
current $J_{i}^{a}$. In the following we use the gauge with vector
potential $\mathbf{A}(t)=\mathbf{A}_{0}e^{-i\omega t}$, $\mathbf{E}%
=-c^{-1}(\partial \mathbf{A}/\partial t)$, and at the end take the limit $%
\omega \to 0$ in the calculated response function \cite{agd}.

Using Eqs.~(\ref{Hamiltonian1}) and (\ref{jia}) one can write the
matrix elements of the spin current operator in the basis of
eigenfunctions of $\hat{H}_{0}$ as
\begin{equation}
\left\langle \overline{\mathbf{k}}|\hat{\jmath}_{i}^{z}|\overline{\mathbf{%
k^{\prime }}}\right\rangle =\frac{\delta _{\overline{\mathbf{kk^{\prime}}}}}{2m}\left(
k_{i}-\frac{eA_{i}}{c}\right) \sigma _{z},
\label{jiz}
\end{equation}
where  $\overline{\mathbf{k}}$ includes the electron momentum
$\mathbf{k}$ and  spin component $\sigma_{z}$. We note that the $z$-component of
spin current, $\hat{\jmath}_{i}^{z}$,  does not contain any
anomalous part explicitly dependent on the SO coupling.

It is convenient to decompose the Hamiltonian $\hat{H}$ into two terms, $%
\hat{H}=\hat{H}_{A=0}+\hat{H}_{A}$, where $\hat{H}_{A=0}$
corresponds to vanishing vector potential
($\mathbf{A}=\mathbf{0}$), while $\hat{H}_{A}$
appears at nonzero $\mathbf{A}$. Matrix elements of Hamiltonian $\hat{H}%
_{A=0}$ are
\begin{eqnarray}
&&\left\langle \overline{\mathbf{k}}|\hat{H}_{A=0}|\overline{\mathbf{k^{\prime
}}}\right\rangle  =\frac{k^{2}}{2m}\delta _{\overline{\mathbf{kk}^{\prime }%
}}+\hat{V}_{\overline{\mathbf{kk^{\prime }}}}, \\
&&\hat{V}_{\overline{\mathbf{kk^{\prime }}}} =\frac{\lambda _{\mathbf{%
kk^{\prime }}}}{2}\left[ \sigma _{x}(k_{y}+k_{y}^{\prime })-\sigma
_{y}(k_{x}+k_{x}^{\prime })\right],
\label{Vkk}
\end{eqnarray}
where $\lambda _{\mathbf{kk^{\prime }}}$ is the Fourier component
of the random
Rashba field. In turn, matrix elements of the $\mathbf{A-}$dependent term, $%
\hat{H}_{A}$, have the form
\begin{eqnarray}
\left\langle \overline{\mathbf{k}}|\hat{H}_{A}|\overline{\mathbf{k^{\prime }}%
}\right\rangle  &=&-\frac{e\mathbf{k}\cdot \mathbf{A}}{mc}\,\delta _{%
\overline{\mathbf{kk}^{\prime }}}+\frac{e^{2}A^{2}}{2mc^{2}}\,\delta _{%
\overline{\mathbf{kk}^{\prime }}}+\hat{W}_{\overline{\mathbf{kk^{\prime }}}},
\label{HA} \\
\hat{W}_{\overline{\mathbf{kk^{\prime }}}} &=&-\frac{e\lambda _{\mathbf{%
kk^{\prime }}}}{c}\left( \sigma _{x}A_{y}-\sigma _{y}A_{x}\right).
\label{Wkk}
\end{eqnarray}
In the linear response regime, only the first and third terms on
the right-hand side of Eq.~(\ref{HA}) are relevant. The third term
clearly demonstrates the coupling of  electric field to electron
spin {\it via} the Fourier component of the random Rashba field.
These two terms can be associated with two different
electromagnetic vertices in the Feynman diagrams for system's
conductivity: the first one leads to the conventional conductivity
while the third one to the spin conductivity.

\begin{figure}[h]
\centering
\includegraphics[width=0.35\textwidth]{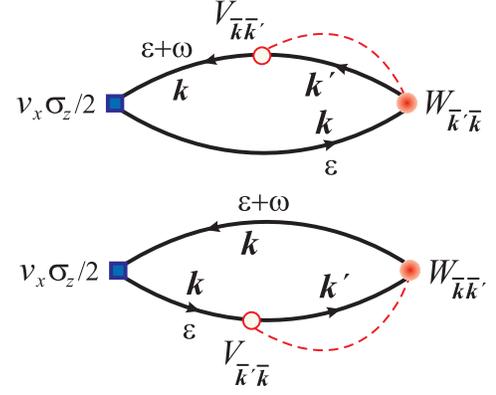}
\caption{The Feynman diagrams leading to nonvanishing
contributions to spin current. Here the left vertex (filled square) corresponds to
the spin current operator, the right vertex (filled circle) is the external field
perturbation in Eq.(\ref{Wkk}), and the white circle is the matrix element of
spin-orbit coupling in Eq.(\ref{Vkk}). Upon averaging
over disorder, the dashed line becomes the Fourier component of
the correlator of random Rashba SO interaction $C({\mathbf
k}-{\mathbf k}^{\prime})$.} \label{fig:int}
\end{figure}

To calculate the spin Hall conductivity we apply the conventional Kubo
formalism \cite{agd} using   the retarded and advanced Green's functions $%
\widehat{G}_{\mathbf{k}}^{R,A}=\hat{I}G_{\mathbf{k}}^{R,A},$ taken
in the vicinity of the Fermi level,
\begin{equation}\label{GRA}
G_{\mathbf{k}}^{R,A}=\frac{1}{\varepsilon_{\mathbf{k}}-\varepsilon_{F}\pm i/2\tau },
\end{equation}
where $\hat{I}$ is the $2\times 2$ unit matrix, $\varepsilon _{\mathbf{k}%
}=k^{2}/2m,$ $\varepsilon_{F}$ is the Fermi energy, and $\tau $ is
the total momentum relaxation time including scattering from
impurities ($\tau_0$) and scattering by spin-dependent Rashba
potential ($\tau_{\rm SO}$), $1/\tau =1/\tau_0+1/\tau_{\rm SO}$.
Since the SO coupling vanishes on the average, the Green function
(\ref{GRA}) keeps exactly the diagonal form in the spin subspace.
The linear spin Hall conductivity is represented by the sum of two
Feynman diagrams in Fig.1. We neglect ladder corrections for spin
current vertex since we assume isotropic scattering by impurities.
On the other hand, corrections to the vertices corresponding to
the random spin-orbit coupling are small by the parameter
$R/\ell\ll 1$, where $R$ is a characteristic length of the
fluctuations in the Rashba interaction, and $\ell $ is the
electron mean free path. This situation is completely different
from the case of constant SO interaction. Since the spin current
operator does not include any anomalous term, there are no
diagrams with the corresponding vertices including the random
Rashba field. In other words, the spin Hall effect is due to a
spin-dependent correction to the distribution function  only, as
will be explicitly verified later in the text by considering the
kinetic equation for the density matrix
\cite{tarasenko06,denmatrix}.

We assume that the field is oriented along the $y$-axis
($A_{x}=0,\;A_{y}\neq 0$) and calculate the spin-Hall conductivity
$\sigma _{\rm sH}$ defined as $J_{x}^{z}=\sigma_{\rm sH}E_{y}$.
One can easily verify that all other components of the spin
current are equal to zero. Calculating the diagrams, taking
the trace in the spin subspace, and integrating over the electron
energy $\varepsilon$, we obtain in the static limit $\omega \to
0$,
\begin{eqnarray}
&&\sigma_{\rm sH}=\frac{ie}{2\pi m}\times \label{sigmasH1}\\
&&\sum_{\mathbf{kq}}k_{x}\left(
k_{x}+k_{x}^{\prime }\right) C(q)\,G_{\mathbf{k}}^{R}\left( G_{\mathbf{k}-%
\mathbf{q}}^{R}-G_{\mathbf{k}-\mathbf{q}}^{A}\right) G_{\mathbf{k}}^{A},\nonumber
\end{eqnarray}
where  $C(\mathbf {q})=C(q)$ for an isotropic system. We consider
the experimentally relevant case of weak SO coupling, where
$\tau_{0}\ll\tau_{\rm SO}$ and, therefore, $\tau$ is very close to $\tau_{0}$.
For the states close to the Fermi surface, the difference
$G_{\mathbf{k}-\mathbf{q}}^{R}-G_{\mathbf{k}-\mathbf{q}}^{A}=2i{\rm Im}G_{%
\mathbf{k}-\mathbf{q}}^{R}$ can be presented in the form
\begin{equation}
G_{\mathbf{k}-\mathbf{q}}^{R}-G_{\mathbf{k}-\mathbf{q}}^{A}
=-2\pi i\delta \left(\varepsilon_{F}-\varepsilon _{\mathbf{k%
}-\mathbf{q}}\right),
\end{equation}
which reflects the energy conservation. By using the resulting
identity,
\begin{eqnarray}
&&\delta \left( \left( k^{2}-\left( \mathbf{k}-\mathbf{q}\right) ^{2}\right)
/2m\right) = \label{delta}\\
&&\quad\frac{2m}{q\sqrt{4k^{2}-q^{2}}}\left[ \delta \left( \theta
-\theta _{1}\right) +\delta \left( \theta -\theta _{2}\right) \right],\nonumber
\end{eqnarray}
with  $\theta $ denoting  the angle between $\mathbf{k}$ and
$\mathbf{q}$, and $\theta_{1,2}$ being two solutions of
$\cos\theta_{1,2}=q/2k$, we arrive upon integrating over $\mathbf{k}$
at
\begin{equation}
\sigma_{\rm sH}=\frac{em\tau }{4\pi ^{2}}\,\int_{0}^{2k_{F}}C(q)\,\sqrt{%
4k_{F}^{2}-q^{2}}\,dq,
\label{sigmasHresult}
\end{equation}
where $k_{F}$ is the Fermi momentum. Taking into account
the formula for the spin relaxation time
$\tau_{s}$ due to random SO coupling, derived in Ref.\cite{dugaev09}
for $\ell\gg R$:
\begin{equation}
\frac{1}{\tau _{s}}=\frac{m}{\pi }\int_{0}^{2k_{F}}C(q)\,\sqrt{%
4k_{F}^{2}-q^{2}}\,dq,
\label{tauso}
\end{equation}
we obtain
\begin{equation}
\sigma _{\rm sH}=\frac{e}{4\pi }\,\frac{\tau }{\tau _{s}}.
\label{sigmasHfinal}
\end{equation}
Equation (\ref{sigmasHfinal}) is our main result. It shows that $\sigma _{\rm sH}$
is non-universal and depends on both the disorder due to
impurities and random SO coupling. However, it is not zero under
very general assumptions of our model, which is qualitatively
different from $\sigma_{\rm sH}=0$ for the uniform Rashba coupling
\cite{dimitrova05}. Of course, our finite $\sigma_{\rm sH}$ is not
in contradiction to the result of Ref. \cite{dimitrova05}  since,
in contrast to Ref. \cite{dimitrova05}, the SO coupling is disordered here.

The above result can also be obtained with the kinetic equation
for random Rashba SO interaction \cite{tarasenko06,dugaev09},
which allows a better insight into the problem. The kinetic
equation for the density
matrix $\hat{\rho}_{\mathbf{k}}$ includes the usual field-dependent term $e%
\mathbf{E\cdot }\partial \hat{\rho}_{0\mathbf{k}}/\partial
\mathbf{k}$, which is responsible for the conductivity. Here the
unperturbed density matrix is
$\hat{\rho}_{0\mathbf{k}}=\hat{I}f_{0}(\varepsilon _{k})$, where
$f_{0}(\varepsilon )$ is the Fermi-Dirac distribution function.
Another source of the perturbation in the electron distribution
under external field $\mathbf{E}$ is due to the spin-dependent
scattering associated with the fluctuating Rashba SO interaction,
as given by the third term in the right-hand-side of Eq.(\ref{HA}). As discussed above,
the electric field produces a random field acting on electron
spin, which can be treated in the collision integral.

Using the matrix elements in Eqs. (\ref{Vkk}), (\ref{Wkk}), and
assuming that the perturbation due to SO interaction is small, we
find the following expression for the collision integral:
\begin{eqnarray}
&&\mathrm{St}\,\hat{\rho}_{\mathbf{k}}
=\frac{1}{\tau }\left( \hat{\rho}_{0\mathbf{k}}
-\hat{\rho}_{\mathbf{k}}\right) + \mathrm{St}^{[E]}\hat{\rho}_{\mathbf{k}},\label{St}\\
&&\mathrm{St}^{[E]}\hat{\rho}_{\mathbf{k}}=2\pi\sum_{\mathbf{k^{\prime }}}
\left(\hat{V}_{\overline{\mathbf{kk^{\prime }}}}
\hat{W}_{\overline{\mathbf{k^{\prime }k}}}+
\hat{W}_{\overline{\mathbf{kk^{\prime }}}}\hat{V}_{\overline{\mathbf{k^{\prime }k}}}\right)\times \nonumber\\
&& \quad \left( \hat{\rho}_{0\mathbf{k^{\prime }}}-\hat{\rho}_{0\mathbf{k}}\right)
\delta (\varepsilon_{\mathbf{k}}-\varepsilon _{\mathbf{k^{\prime }}%
}+\omega), \label{StE}
\end{eqnarray}
where $\mathrm{St}^{[E]}\hat{\rho}_{\mathbf{k}}$ is the
contribution from the random Rashba field. Since we consider the
linear response to $E_{y}$, in the last term we take the
equilibrium density matrix, with $\hat{\rho}_{0\mathbf{k^{\prime }}}-\hat{%
\rho}_{0\mathbf{k}}=\omega \, \partial f_{0}(\varepsilon )/\partial \varepsilon
$. Upon the
averaging over the SO disorder, we obtain,
\begin{equation}\label{mat:el}
\hat{V}_{\overline{\mathbf{kk^{\prime }}}}
\hat{W}_{\overline{\mathbf{k^{\prime }k}}}+
\hat{W}_{\overline{\mathbf{kk^{\prime }}}}\hat{V}_{\overline{\mathbf{k^{\prime }k}}}
=-\frac{e}{\omega }%
C(q)\left( k_{x}+k_{x}^{\prime }\right) \sigma _{z}E_{y}.
\end{equation}
This term is the driving force for the spin current, as shown in Fig.2. The
corresponding contribution to the collision integral can be
obtained with Eq.(\ref{delta}) as
\begin{equation}
\mathrm{St}_{k}^{[E]}=-e\frac{\partial f_{0}}{\partial \varepsilon }\frac{1%
}{k}\frac{1}{\tau _{s}}\left( \frac{k_{x}}{k}\sigma _{z}\right) E_{y}.
\end{equation}
We present the density matrix as $\hat{\rho}_{\mathbf{k}}=\hat{\rho}_{0\mathbf{k}}
+\delta\hat{\rho}_{\mathbf{k}}+S_{\mathbf{k}}\sigma _{z}$ and
with Eqs.~(\ref{St}),(\ref{StE}), find for the steady state:
\begin{equation}
S_{\mathbf{k}}=-e\frac{\partial f_{0}}{\partial \varepsilon }\frac{k_{x}}{%
k^{2}}\frac{\tau }{\tau _{s}}E_{y},
\end{equation}
describing spin split of the Fermi surface, corresponding to Fig.2.
In the stationary state the spin-dependent force due to the external field
and random spin-orbit coupling in Eq.(\ref{mat:el}) is balanced by
the friction force  due to the disorder, proportional to $1/\tau$.

Having found the distribution function, we can calculate the spin
current. First, we calculate in the spin space $\mathrm{Tr}\rho
_{\mathbf{k}}j_{x}^{z}$, with $j_{x}^{z}=k_{x}\sigma _{z}/2m$ and obtain:
\begin{equation}
\mathrm{Tr}\rho _{\mathbf{k}}j_{x}^{z}=-e\frac{\partial
f_{0}}{\partial \varepsilon }\frac{\tau }{\tau
_{s}}\frac{k_{x}^{2}}{mk_{F}^{2}}E_{y}.
\end{equation}
The total spin current is then equal
\begin{equation}
J_{x}^{z}=\int \mathrm{Tr}\, \rho
_{\mathbf{k}}j_{x}^{z}\frac{d^{2}k}{\left( 2\pi \right)
^{2}}=\frac{e}{4\pi }\frac{\tau }{\tau _{s}}E_{y}.
\end{equation}
This result leads to $\sigma _{\rm sH}$ equivalent to
Eq.(\ref{sigmasHfinal}).

\begin{figure}[h]
\centering
\includegraphics[width=0.35\textwidth]{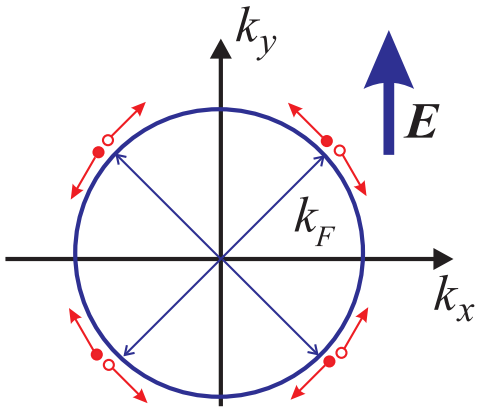}
    \caption{Preferable spin-dependent scattering direction by the effective
     potential in left-hand-side of Eq.(\ref{mat:el}), shown by
     horizontal arrows attached to the circles. White and filled circles correspond to
     spin components $\sigma_{z}=1$ and $\sigma_{z}=-1$, respectively.  As a result,
     the Fermi line becomes spin split with the preferable concentration of
     spin-up electrons at $k_x>0$ and spin-down ones at $k_x<0$, leading to the SHE.}
    \label{fig:fermi_sur}
\end{figure}

To consider an example, we assume the following form of the correlator $C(q)$
\cite{sherman03,dugaev09}:
\begin{equation}
C(q)=2\pi \left\langle \lambda ^{2}\right\rangle R^{2}e^{-qR},
\label{9}
\end{equation}
achieved by doping quantum wells with charged impurities. In the
semiclassical limit of long-range correlations, $k_{F}R\gg 1$, the
integral in Eq.~(\ref{tauso}) becomes
\begin{equation}
\frac{1}{\tau _{s}}=2k_{F}\frac{m}{\pi }\int\limits_{0}^{\infty
}C(q)\,dq=4m\left\langle \lambda ^{2}\right\rangle k_{F}R,
\label{10}
\end{equation}
and the resulting spin Hall conductivity is
\begin{equation}
\sigma _{\rm sH}=\frac{e}{\pi }\,m\tau \left\langle \lambda ^{2}\right\rangle
k_{F}R.  \label{11}
\end{equation}
When $k_{F}R\ll 1$, the system is always in the dirty limit of
very long SO coupling-determined relaxation times. As a result,
the spin Hall conductivity is suppressed and tends to zero as
$\left(k_{F}R\right)^{2}$ with decreasing $k_{F}R$. This is a general
feature of the finite-range correlators (similar to that in
Eq.(\ref{9})), the effect of which vanishes due to fast
oscillations of the Rashba parameter on the spatial scale of the
electron wavelength.

Now we can qualitatively discuss the clean limit  $\tau _{0}\gg
\tau_{\rm SO}.$ Here  the spin conductivity is finite and
does not depend on the magnitude of fluctuating Rashba field since both the relaxation
rate and gain due to the external field are proportional to $\left\langle \lambda
^{2}\right\rangle $. However, the clean limit requires a separate analysis of the
relaxation timescales, which will be considered elsewhere.

In conclusion, we have shown that the random Rashba spin-orbit
interaction can generate spin Hall effect, even in the presence of
impurities. This behavior is distinct from that found for
spatially uniform Rashba interaction, where in the limit of small
impurity concentration, the potential scattering by impurities
totally suppresses the spin Hall effect. In contrast to Ref.
[\onlinecite{Rashba04}], where it was found that for the linear
Rashba coupling this suppression is a result of a sum rule for
spin conductivity, here the corresponding rule cannot be
established and the resulting spin current is not suppressed. We
mention that the comparison of conventional and torque-related
\cite{Shi} definitions of spin current shows that the result in
Eq.(\ref{sigmasHfinal}) is independent of definition.

In systems with nonzero spin polarization, arising for instance
due to finite magnetization, the above discussed SHE
is closely related to the anomalous Hall effect. If the
concentrations of spin-up and spin-down electrons are different,
spin separation leads to electric current which gives rise to the
anomalous Hall effect. Since the disorder in spin-orbit coupling
leads to the spin current, it can cause the anomalous Hall effect,
too.

This work was supported by Polish Ministry of Science and Higher
Education as a research project in years 2007 -- 2010 and by 
FCT Grant No.~PTDC/FIS/70843/2006 in Portugal.  E.Y.S.
is supported by the University of Basque Country UPV/EHU grant
GIU07/40, Basque Country Government (grant IT-472-10), and MCINN
of Spain grant FIS2009-12773-C02-01.

\end{document}